\documentclass[conference]{IEEEtran}
\IEEEoverridecommandlockouts
\usepackage{cite}
\usepackage{amsmath,amssymb,amsfonts}
\usepackage{algorithmic}
\usepackage{graphicx}
\usepackage{subfigure}
\usepackage{textcomp}
\usepackage{xcolor}
\usepackage{stfloats}
\usepackage{array}
\usepackage{bm}

\usepackage[justification=centering]{caption}
\def\BibTeX{{\rm B\kern-.05em{\sc i\kern-.025em b}\kern-.08em
		T\kern-.1667em\lower.7ex\hbox{E}\kern-.125emX}}
\begin{document}
	\title{An Improved Equiangular Division Algorithm for SBR based Ray Tracing Channel Modeling}
	
	\author{Yuyang Zhou\textsuperscript{1}, Yinghua Wang\textsuperscript{2}, Yuxiao Li\textsuperscript{1}, Jialing Huang\textsuperscript{1}, Jie Huang\textsuperscript{1,2*}, and Cheng-Xiang Wang\textsuperscript{1,2*}
	\\
	\textsuperscript{1}National Mobile Communications Research Laboratory, School of Information of Science and Engineering, 
	\\Southeast University, Nanjing 210096, China.\\
	\textsuperscript{2}Purple Mountain Laboratories, Nanjing 211111, China.\\
	\textsuperscript{*}Corresponding Authors: Jie Huang, Cheng-Xiang Wang\\
	Email: yyzhouz@126.com, wangyinghua@pmlabs.com.cn, \{yuxli, jlhuang, j\_huang, chxwang\}@seu.edu.cn
	
}

	\maketitle
	\begin{abstract}
		Compared with image method (IM) based ray tracing (RT), shooting and bouncing ray (SBR) method is characterized by fast speed but low accuracy. In this paper, an iterative precise algorithm based on equiangular division is proposed to make rough paths accurate, allowing SBR to calculate exact channel information. Different ray launching methods are compared to obtain a better launching method. By using equiangular division, rays are launched more uniformly from transmitter (Tx) compared with the current equidistant division method. With the proposed iterative precise algorithm, error of angle of departure (AOD) and angle of arrival (AOA) is below 0.01 degree. The relationship between the number of iterations and error reduction is also given. It is illustrated that the proposed method has the same accuracy as IM by comparing the power delay profile (PDP) and angle distribution of paths. This can solve the problem of low accuracy brougth by SBR.

	\end{abstract}
	\section{Introduction}
	RT technology, as a deterministic wireless channel modeling method, is able to simulate the propagation process of each ray through tracking and calculation, so as to accurately predict the propagation results of wireless signals, which satisfies the needs of 6G\cite{b1,b2}. Rays propagate in the environment, reflecting on surfaces, diffracting on wedges, and scattering on rough faces. The power of each ray is therefore calculated \cite{ref1,ref2,ref3}. In addition to predicting the power of the received signal, RT can also provide information such as the pulse time delay of the ray propagation. By calculating the signal amplitude, time delay, and AOA of each multipath signal component, the dispersion of the channel in time, frequency, and space can also be obtained. Compared with the statistical channel model, RT model can accurately reflect the channel characteristics of different scenarios and has great advantages in prediction accuracy, which provides better support for the design of wireless communication systems.
	
	There are two main categories of RT technologies, known as forward RT (represented by SBR) and backward RT (represented by IM) \cite{ref4}. SBR is characterized by high speed but low accuracy, while IM is characterized by high accuracy but slow speed. Also, it is pointed out that the time complexity of the IM algorithm is $O(k^j)$, where $k$ is the total amount of faces and wedges, and $j$ is the sum of the reflection and diffraction orders. Therefore, in the case of complex urban scenes and large number of reflections, IM takes a long time to obtain results \cite{ref4}. Moreover, a hybrid method is proposed \cite{ref5}. It first combines SBR with IM, employing IM to adjust the inaccurate trajectories obtained by SBR, which reserves both the fast speed of SBR and the accuracy of IM.
	
	 Most research on SBR algorithms focused on applying SBR to different scenarios, extending SBR to multiple input multiple output (MIMO), or innovating acceleration algorithms \cite{ref7, refMIMO, refsen}. But few of them improved the accuracy of SBR. Though some work has been done to advance the ray launching method \cite{ref9}, it is limited to a simple scenario with only reflection. A Ray-jumping method is mentioned in \cite{refjump} to reduce the error, but the time cost is rather high with much more rays traced.
	
	This article presents the comparison of different ray launching methods, together with an improved ray launching method based on angle bisector. Also, an analysis of the time cost of improving accuracy in SBR technology is put forward. An iterative algorithm for improving the accuracy of SBR is shown. By relaunching sub-rays around the original ray cone, more precise paths are obtained. The relationship between accuracy improvement and the number of iterations is derived and verified by several scenes. Finally, the results are compared with IM, which is proved to be the most precise RT method, proving accuracy improvement of iterative algorithm.
	
	The remainder of this paper is organized as follows. Section~\uppercase\expandafter{\romannumeral2} introduces a new ray launching method. Section~\uppercase\expandafter{\romannumeral3} explains the steps and expectations of the iterative precise algorithm. Simulation results and analysis are presented inSection~\uppercase\expandafter{\romannumeral4}. Conclusions are drawn in Section~\uppercase\expandafter{\romannumeral5}.

	\section{Ray Launching Methods}	
	\subsection{Equidistant Division}	
\begin{figure}[tb]
		\centering
		\vspace{-2mm}
		\subfigure[]{
			\centering
			\includegraphics[scale=0.2]{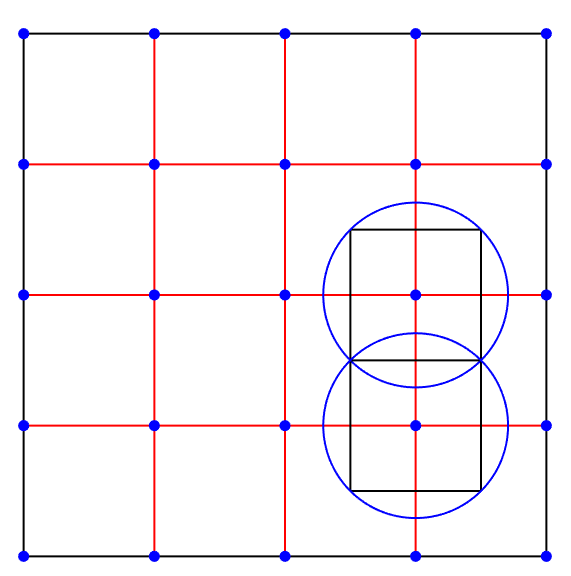}
			
		}
		\subfigure[]{
			\centering
			\includegraphics[scale=0.22]{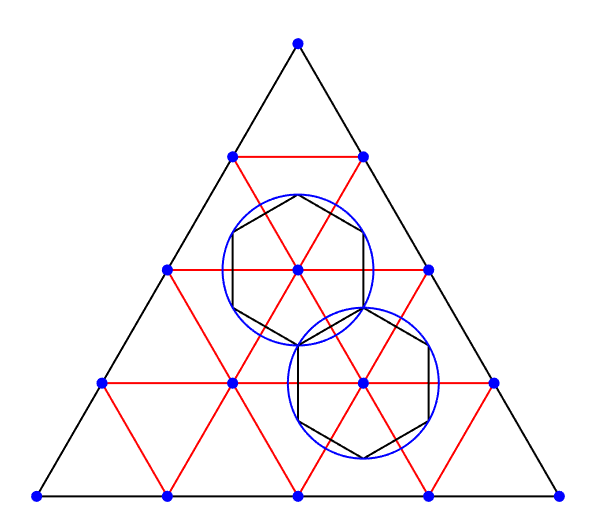}
			
		}
		\setlength{\belowcaptionskip}{-0.5cm}
		\vspace{-2mm}
		\caption{Illustration of division methods, (a) square face division method and (b) triangular face division method, with ray cone coverage area in circles, rays in dots, actual ray cone coverage area in squares (a) and hexagons (b).}
		\vspace{-2mm}
		\label{Fig1}
	\end{figure}

	In SBR, rays are launched on the faces of a regular polyhedron by dividing the edges and connecting division points. Compared with a hexahedron whose faces are squares, a regular polyhedron with triangular faces is more suitable for launching rays. Because the rays transmit in the form of cones, overlapping area is inevitable, which causes double counting. As shown in Fig.~\ref{Fig1}, a square has larger adjacent overlapping area than a triangle, causing the hexahedra inappropriate for ray launching. A dodecahedron is less suitable owing to its pentagon faces which cannot be flattened.

\begin{figure}[b]
		\centering
		\vspace{-10mm}
		\subfigure[]{
			\centering
			\includegraphics[scale=0.26,trim=50 30 50 30,clip]{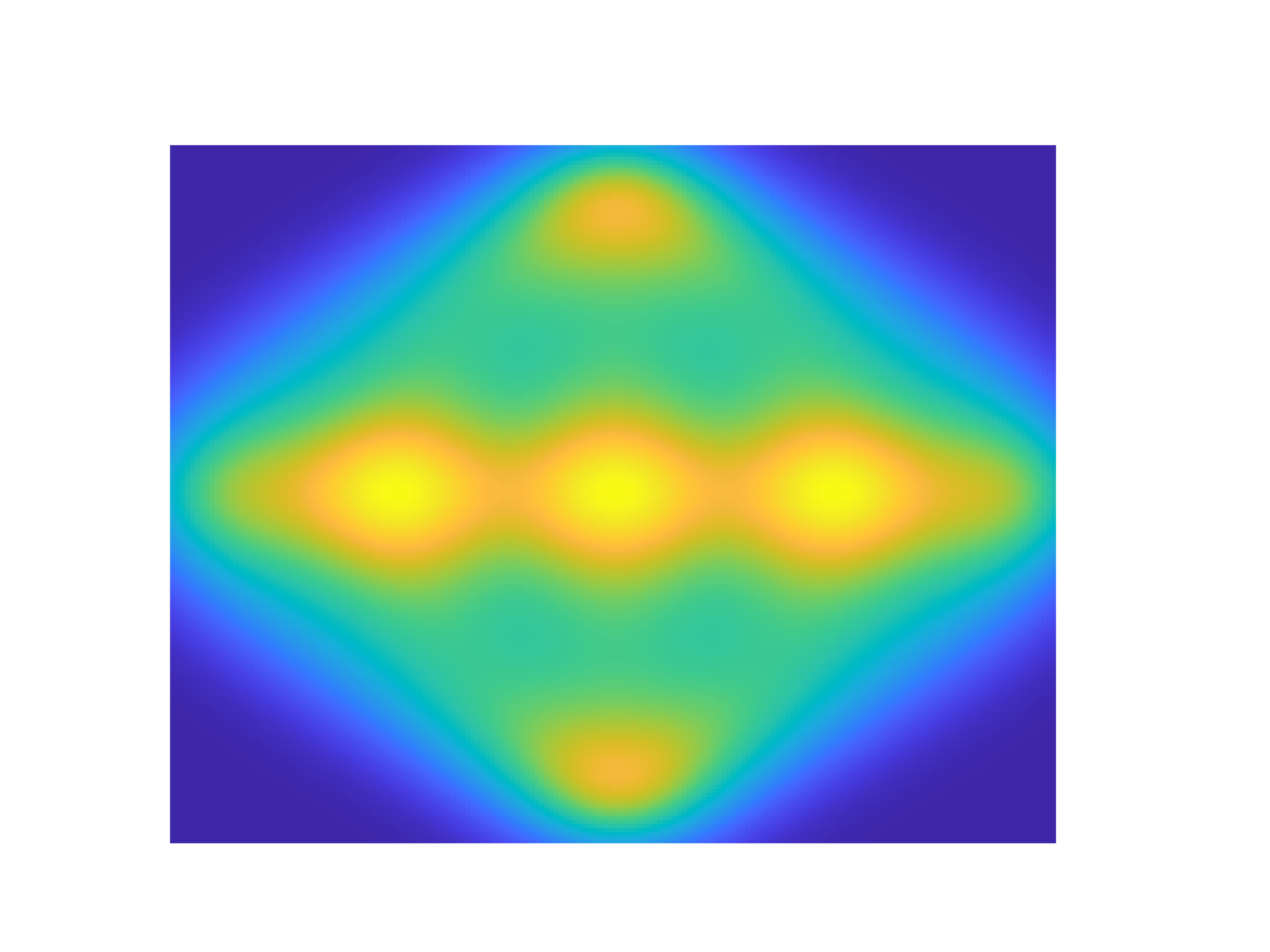}		
			\label{Fig2a}
		}
		\hspace{-6mm}
		\subfigure[]{
			\centering
			\includegraphics[scale=0.26,trim=50 30 50 30,clip]{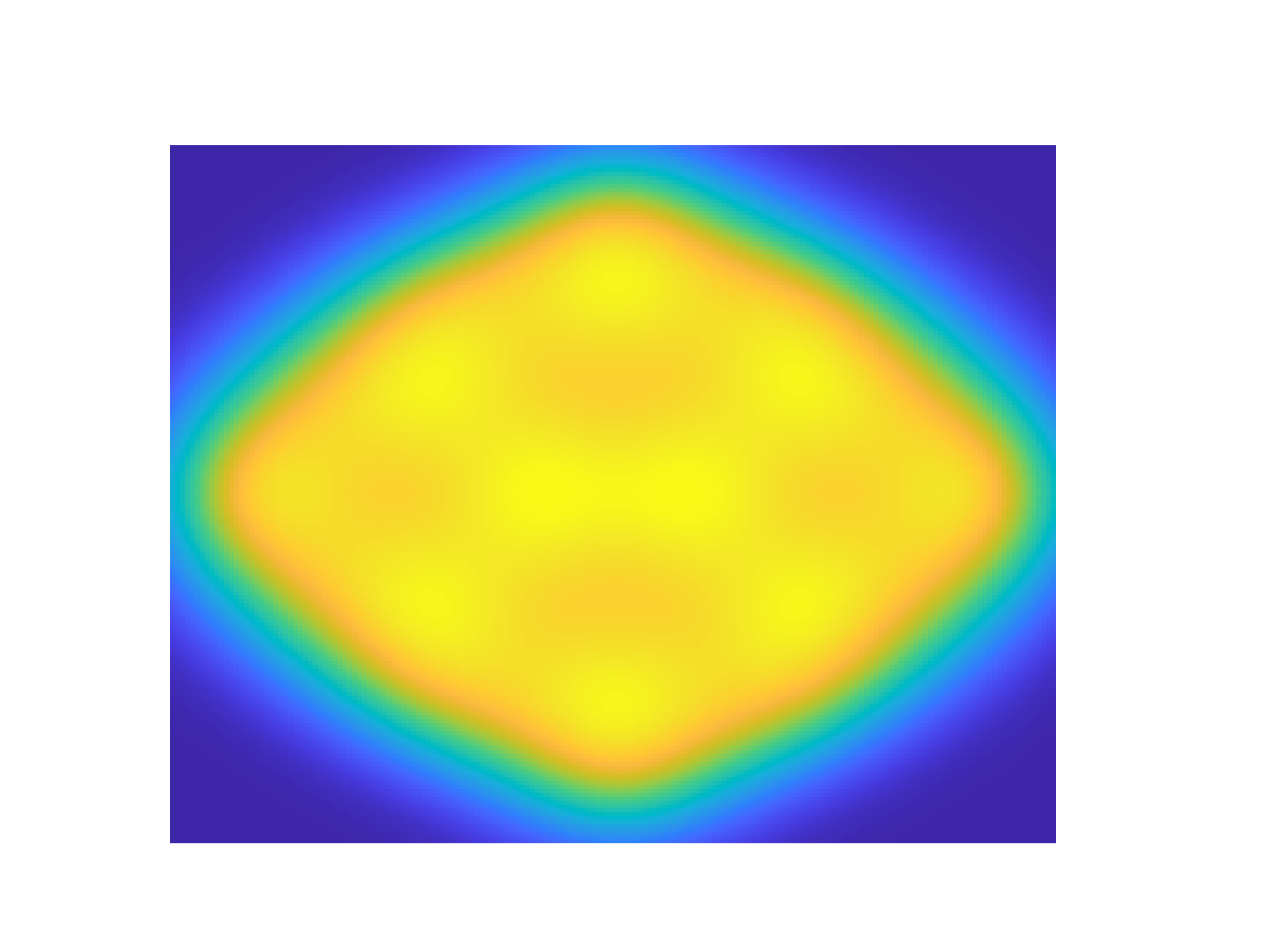}	
			\label{Fig2b}	
		}
		\hspace{-6mm}
		\subfigure[]{
			\centering
			\includegraphics[scale=0.26,trim=50 30 50 30,clip]{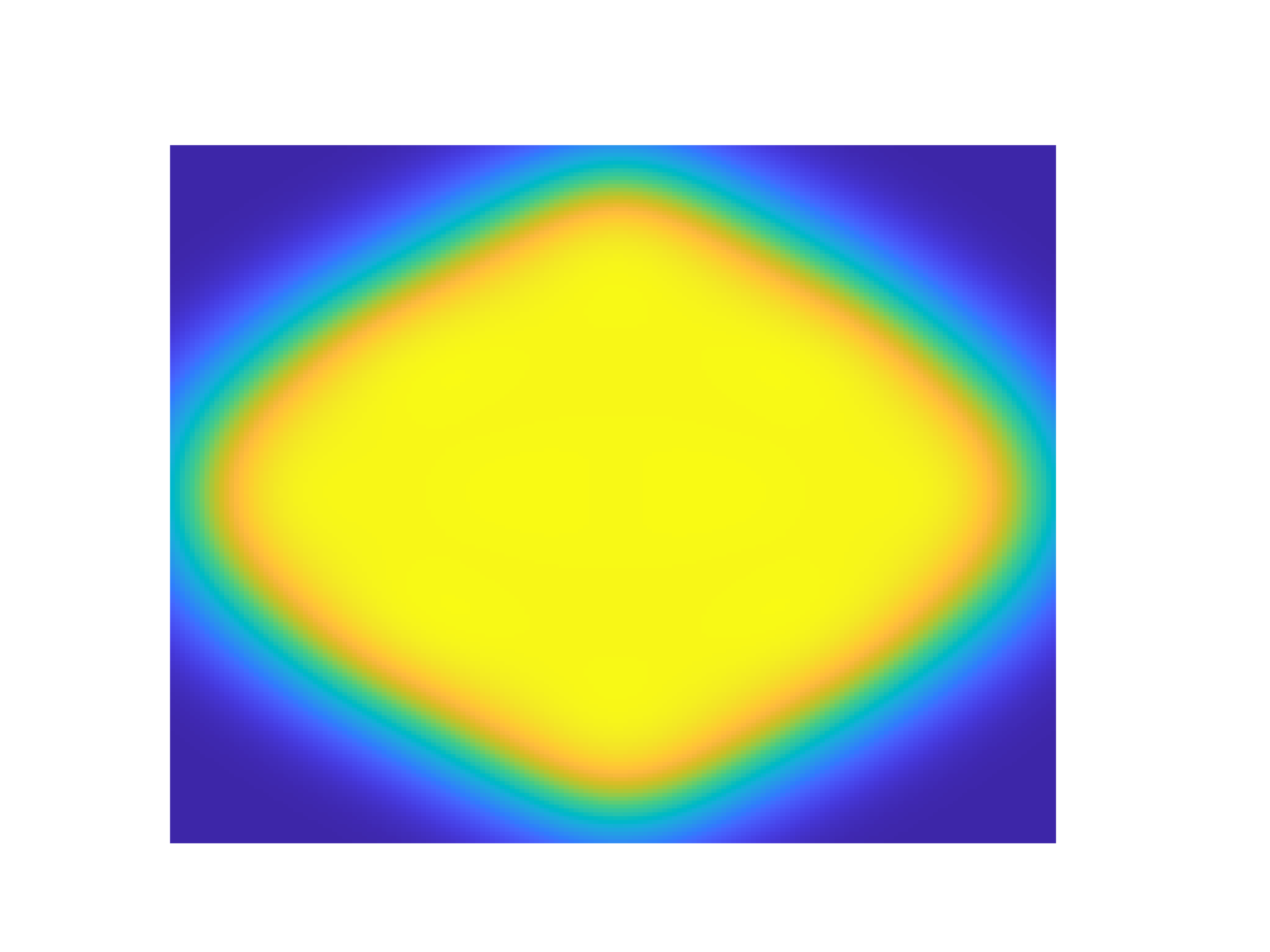}
			\label{Fig2c}		
		}
		\qquad

		\setlength{\belowcaptionskip}{-0.5cm}
		\vspace{-4mm}
		\caption{Density map of launching 4412 rays with equidistant division on (a) an octahedron, (b) an icosahedron, and (c) with equiangular division on an icosahedron.}
		\label{Fig2}
	\end{figure}

	The method in \cite{ref9} generates ray points at equal intervals on the faces of the regular polyhedron and then projects these points onto the sphere, referred to as equidistant division here. In the projection process, the spacing of rays launched near the polygon face center expands more, while that of rays launched near vertices hardly changes, which leads to a smaller ray density at the face center than that at the vertices. Compared with an icosahedron, an octahedron has fewer faces, larger area of triangular faces, and a farther distance to the spherical surface at face centers, so this diffusion effect becomes more obvious, verified in Fig.~\ref{Fig2a} and Fig.~\ref{Fig2b}. A tetrahedron has even fewer faces than an octahedron, so it has the worst performance among regular polyhedrons with triangular faces.
	
	By calculating the azimuth and elevation angles of rays, and drawing a density map, the ray launching density of different ray emission methods can be compared. But it should be noted that the horizontal circular section of a sphere is the longest at the equator and shortest at the poles. The radius of the horizontal circle is $R\cos(pitch)$, where $R$ is radius of the sphere. In order to make a fair comparison, we should use azimuth multiplied by $\cos(pitch)$ as the abscissa of figures, instead of azimuth angle simply. In this way, the sphere can be expanded to a two-dimensional plane, with the area unchanged as $4{\pi}R^2$. The density of points on the sphere suffers little impact in this way, and the visualization of the density map is better due to expansion to the plane.
	
	The color in Fig.~\ref{Fig2a} is less uniform than that in Fig.~\ref{Fig2b}, meaning an icosahedron is more suitable for launching rays. However, there is still a ray density gap between the vertices and the face centers. A new approach enabling ray density much more uniform is proposed below.

	\subsection{Equiangular Division}
	
\begin{figure}[tb]
		\centering
		\subfigure[]{
			\centering
			\includegraphics[scale=0.5,trim=0 0 0 0,clip]{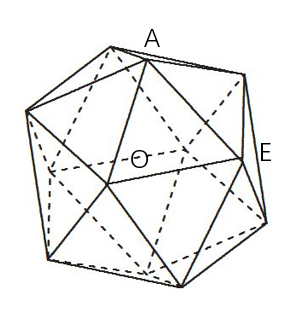}		
			\label{Fig3a}
		}
		\hspace{-6mm}
		\subfigure[]{
			\centering
			\includegraphics[scale=0.35,trim=0 0 0 0,clip]{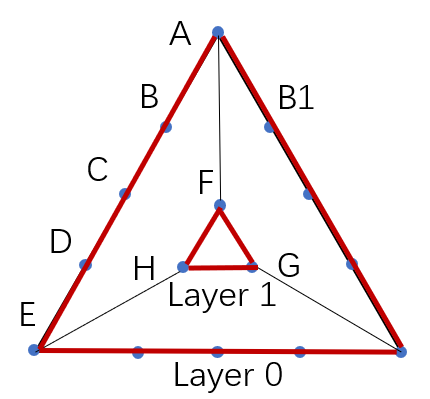}	
			\label{Fig3b}	
		}

		\setlength{\belowcaptionskip}{-0.5cm}
		
		\caption{(a) a regular icosahedron, with A and E adjacent vertices and O the center, (b) iterative steps of launching rays with B, C and D positioned on edge AE.}
		\label{Fig3}
	\end{figure}

	The results of the equidistant division on an icosahedron still have a problem of high ray density at the vertices. This is an inherent problem that comes with equidistant division. Since the rays are finally launched on a spherical surface, and the actual distance is related to the angle between the rays, ray positions should be divided by equal angles in the beginning. Fig.~\ref{Fig3a} is an icosahedron, on which rays are launched with the method in Fig.~\ref{Fig3b}. The equidistant division method satisfies $AB=BC=CD=DE$, while the equiangular division method satisfies $\angle{AOB}=\angle{BOC}=\angle{COD}=\angle{DOE}$, where $O$ is the center of the icosahedron circumscribed sphere. In this way, the angular intervals of ray points on the edge are the same so that ray density is the same. The inner layer is constructed using an iterative method. As shown in Fig.~\ref{Fig3b}, $\overrightarrow{AF}=\overrightarrow{AB}+\overrightarrow{AB_1}$. New vertices ($F, H and G$) on the inner layer are constructed.

	Here, $P^{b}_{a}$ is defined as one ray point launched on one triangular surface of the icosahedron, where $a$ is the layer number and $b$ is the point number on the layer. The outermost layer of the triangular is layer 0. When an inner layer is generated, the number is added by 1, shown in Fig.~\ref{Fig3b}. The ray points are arranged clockwise on the triangle, with $A$ being the first point $P^{0}_{0}$. The edges are divided into $n$ subdivisions. In this way, $P^{0}_{0}$, $P^{n}_{0}$ and $P^{2n}_{0}$ are the vertices of the outermost triangle layer. Ray points are generated by	
	\begin{equation}\label{eq1}
	\angle{P^{b}_{a}OP^{b+1}_{a}}=\frac{\angle{P^{0}_{0}OP^{n-3a}_{0}}}{n-3a}, b=1,2,...,3n-9a-2
	\end{equation}
	with $a=0$, launching $3n$ rays on the outermost layer. 
	
	Then, the vertices of the inner layer $P^{0}_{a+1}$, $P^{n-3a-3}_{a+1}$ and $P^{2n-6a-6}_{a+1}$ are generated by its outer layer
	\begin{equation}
		\begin{aligned}
			\overrightarrow{P^{0}_{a}P^{0}_{a+1}}=\overrightarrow{P^{0}_{a}P^{1}_{a}}+\overrightarrow{P^{0}_{a}P^{3n-9a-2}_{a}}\\
			\overrightarrow{P^{n-3a}_{a}P^{n-3a-3}_{a+1}}=\overrightarrow{P^{n-3a}_{a}P^{n-3a+1}_{a}}+\overrightarrow{P^{n-3a}_{a}P^{n-3a-1}_{a}}\\
			\overrightarrow{P^{2n-6a}_{a}P^{2n-6a-6}_{a+1}}=\overrightarrow{P^{2n-6a}_{a}P^{2n-6a+1}_{a}}+\overrightarrow{P^{2n-6a}_{a}P^{2n-6a-1}_{a}}.\\
		\end{aligned}
	\end{equation}	
	The $3n-9a$ ray points on the layer $a$ are generated by (\ref{eq1}).

	There are three possible situations on the innermost layer, with the number and location of the innermost rays being none; one ray at the center of the face; three rays generated from the outer layer, depending on $n$. Each face of an icosahedron is divided in this way to launch rays.
	
	The numbers of rays launched by both methods are the same. It is shown in Fig.~\ref{Fig2b} and Fig.~\ref{Fig2c} that equiangular division enables the rays to distribute more uniformly after projection. So, a consistent angle can be adopted at the receiving end. Also, this uniform division is the basis for the subsequent precise algorithm.

	\section{Iterative Precise Algorithm}
	\subsection{Tracing Rays}
\begin{figure}[tb]
		\centering
		\vspace{-0mm}
		\includegraphics[scale=0.25,trim=0 40 0 0,clip]{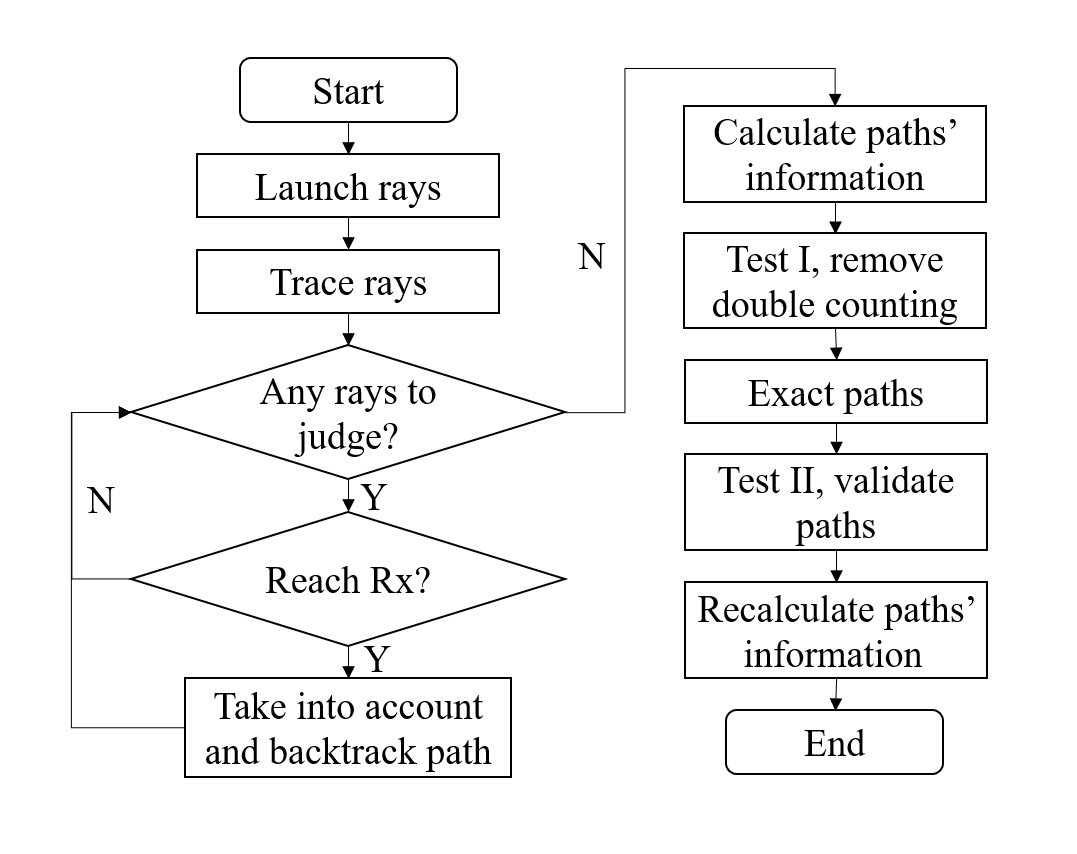}
		\setlength{\belowcaptionskip}{-0.5cm}
		\caption{Flowchart of the algorithm.}
		\label{Fig4}
	\end{figure}

	The flowchart of SBR in this paper is in Fig.~\ref{Fig4}. Propagation mechanisms of ray-tracing include reflection and diffraction.

	There are two path test parts. The first step is for removing double counting. The transmitting end inevitably has overlapping areas, so it is necessary to exclude repeated rays at the receiving end. The method of exclusion is excluding the one with a larger error angle if the two paths pass through exactly the same faces or wedges. Here the error angle is defined as
	\begin{equation}\label{eq3}
		\theta_{error}=\arctan(\frac{d_{Rx-ray}}{d_{ray}})
	\end{equation}
	where $d_{Rx-ray}$ is the distance between Rx and the ray that satisfies the reception judgment, and $d_{ray}$ is the propagation distance of the ray. Error angle is the angle between departure direction of accurate path and the obtained path from Tx.

	The second test can judge validity of the paths, ensuring the paths collide with the faces or wedges. Also, because the rays of SBR are cones, there may be some redundant paths. After precision, the collision points of these paths are actually outside the faces or wedges, so these paths should be excluded.

	\subsection{Precise Algorithm}
	The path results obtained by the SBR algorithm are inaccurate because the rays are emitted in the form of cones. Error angle is used to compare the error of each path which is defined in (\ref{eq3}). Since the condition for the path to be accepted is that Rx is inside the ray cone, we have $\theta_{error}\leqslant\theta_{launch}$.
	
	Reducing the cone angle of ray emission can effectively reduce the error, and increasing the number of emitted rays can reduce the cone angle. If the emitting end takes $n$ equal parts to launch rays on the regular icosahedron, after some derivation, the total number of rays is $N_{rays}=10n^2+2$, and cone angle is
	\begin{equation}\label{eq8}
		\theta_{launch}=\frac{\theta_{adjacent}}{\sqrt{3}}=\frac{\theta_{0}}{\sqrt{3}n}
	\end{equation}
	where $\theta_{0}$ is the angle connecting the adjacent vertices of the icosahedron to the center, as $\angle{AOE}$ in Fig.~\ref{Fig3b}, equaling to $63.4349^{\circ}$. So, the relationship between error angle and number of rays is
	
	\begin{equation}
		\theta_{error}\leqslant\frac{\theta_{0}}{\sqrt{3}\sqrt{\frac{N_{rays}-2}{10}}}=\frac{115.8158^{\circ}}{\sqrt{N_{rays}-2}}.
	\end{equation}	
	To reduce the error, it is necessary to greatly increase the number of emitted rays. This results in many invalid rays, which do not constitute paths but need to be calculated. Time cost increases greatly and error decreases insignificantly if error angles are reduced by increasing the number of rays.

	When the number of rays is $4412$, all paths can be traced for most scenes. In order to improve accuracy, the paths should be processed, rather than many useless rays to be launched. An exact algorithm that takes very little time is proposed below.
	
\begin{figure}[tb]
		\centering
		\subfigure[]{
			\centering
			\includegraphics[scale=0.44,trim=0 0 0 0,clip]{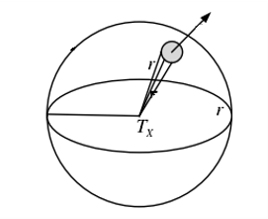}		
			\label{Fig5a}
		}
		\hspace{-8mm}
		\subfigure[]{
			\centering
			\includegraphics[scale=0.32,trim=80 40 50 40,clip]{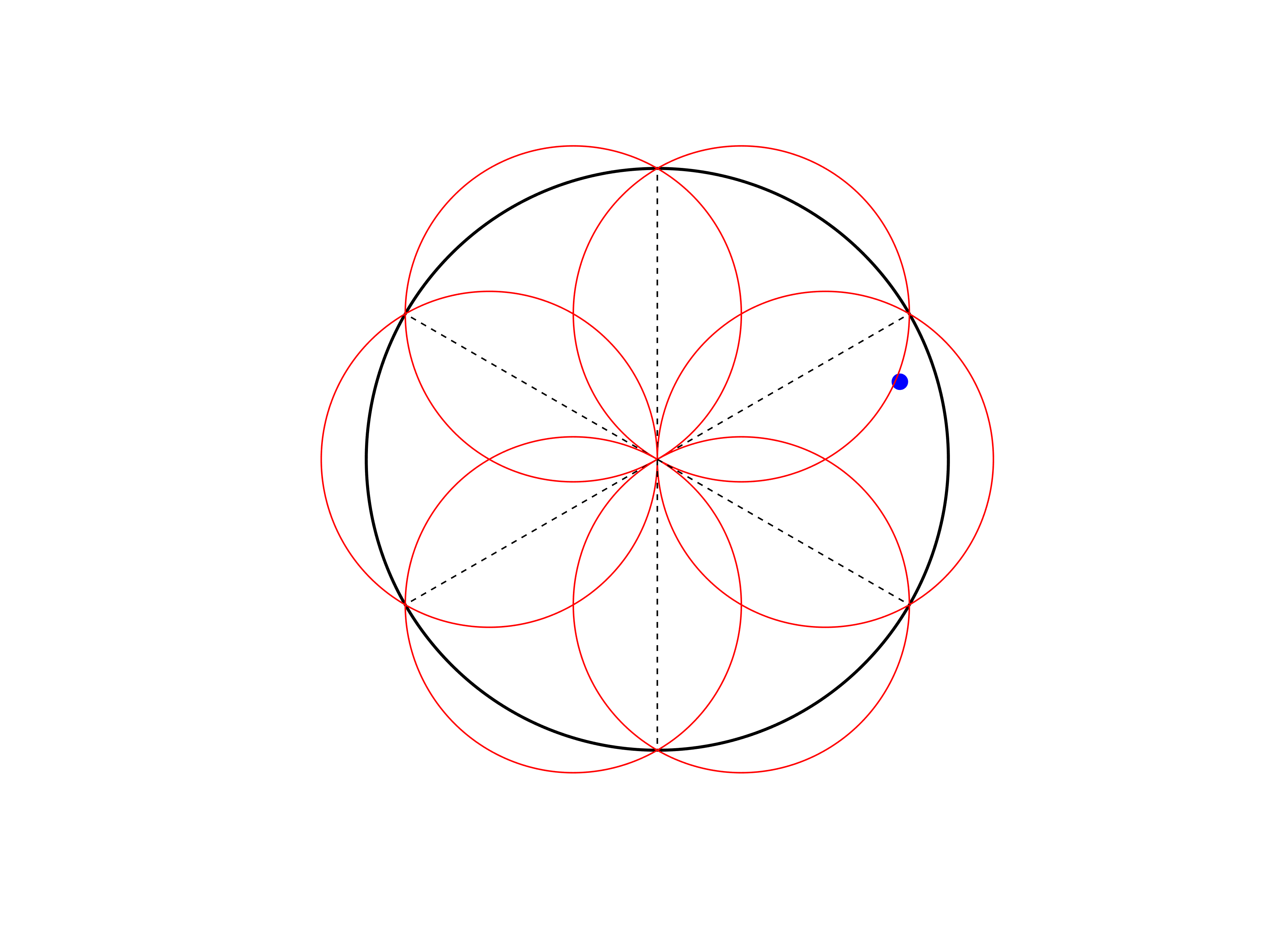}	
			\label{Fig5b}	
		}
		\qquad		
		\subfigure[]{
			\centering
			\includegraphics[scale=0.32,trim=80 40 80 40,clip]{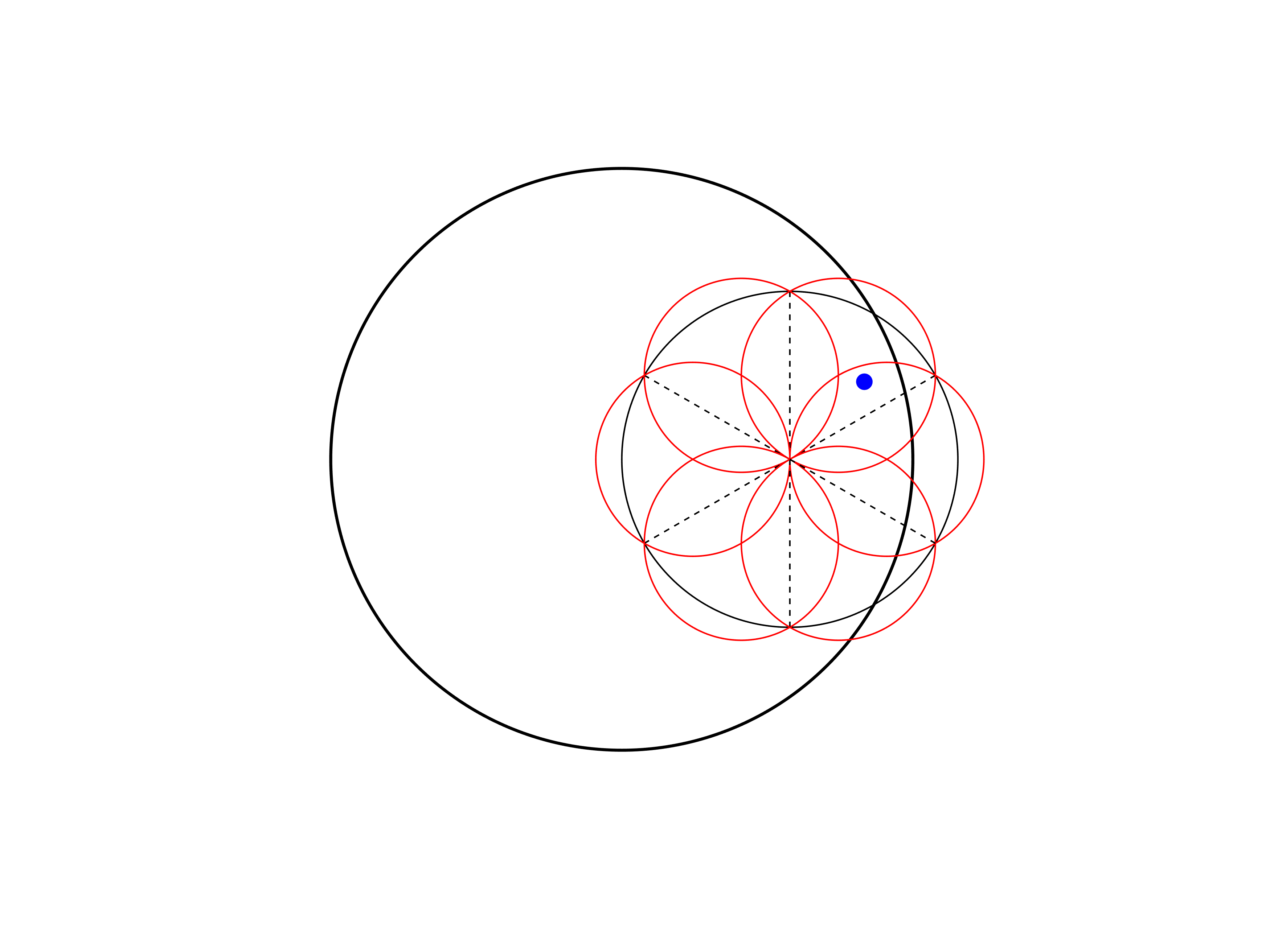}
			\label{Fig5c}		
		}
		\hspace{-6mm}
		\subfigure[]{
			\centering
			\includegraphics[scale=0.32,trim=80 40 80 40,clip]{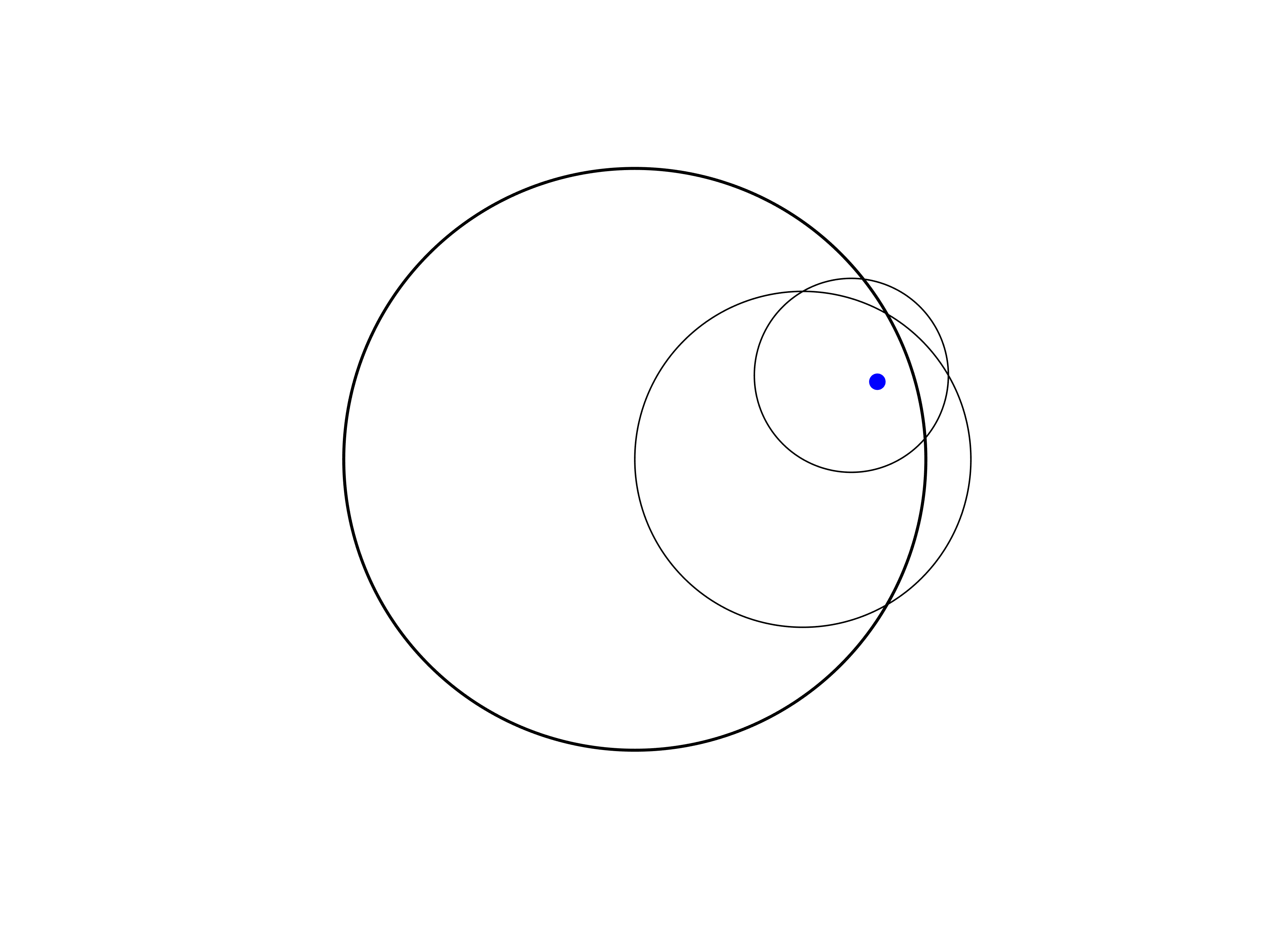}
			\label{Fig5d}		
			
		}
		\setlength{\belowcaptionskip}{-0.5cm}
		
		\caption{(a) ray cone consisting path launched from Tx, (b) first run of the method in bottom view of the ray cone, with original ray cone in large circle, sub-rays in small circles, and accurate ray direction in dot, (c) second run of the algorithm, and (d) more precise ray cone in smallest circle.}
		\label{Fig5}
	\end{figure}

	As shown in Fig.~\ref{Fig5b}, the large circle represents the initial ray cone, and this area is divided into six small areas, represented by the smaller circles. Rays are emitted in these six sub-regions respectively, then several paths are obtained. The error angles of these paths are calculated. The path with the smallest error angle is reserved, and six sub-rays are emitted again within its ray cone, as shown in Fig.~\ref{Fig5c}. Finally, the ray cones iterate in this way until the error angle satisfies the requirements or times of run reach the set number of iterations.

	The directions of all diffraction paths are recorded in order to reduce the time consumption of the precise algorithm. The sub-rays diffract according to the recorded direction, and several rays are emitted to both sides of the diffraction cone. If all the diffraction rays of the six sub-rays are recalculated, there will be a lot of invalid rays, which do not consist of paths. This will waste much time.	
	
	Using this exact algorithm, the ray cone angle shrinks to $1/\sqrt{3}$ each iteration, so the error angle for the $ith$ iteration is	
	\begin{equation}\label{eq10}
		\theta_{error_i}\leqslant\frac{\theta_{launch}}{3^{i/2}}=\frac{63.4349^{\circ}}{3^{(i+1)/2}\cdot{n}}.
	\end{equation}
	In this way, 10 iterations can lead to 24dB reduction in error of angle, distance and power. Besides, exponential error reduction can be achieved with a linear time penalty.

	\section{Results and Analysis}
\begin{figure}[t]
	\centering
	\vspace{0mm}
	\subfigure[]{
		\centering
		\includegraphics[scale=0.5,trim=90 0 70 40,clip]{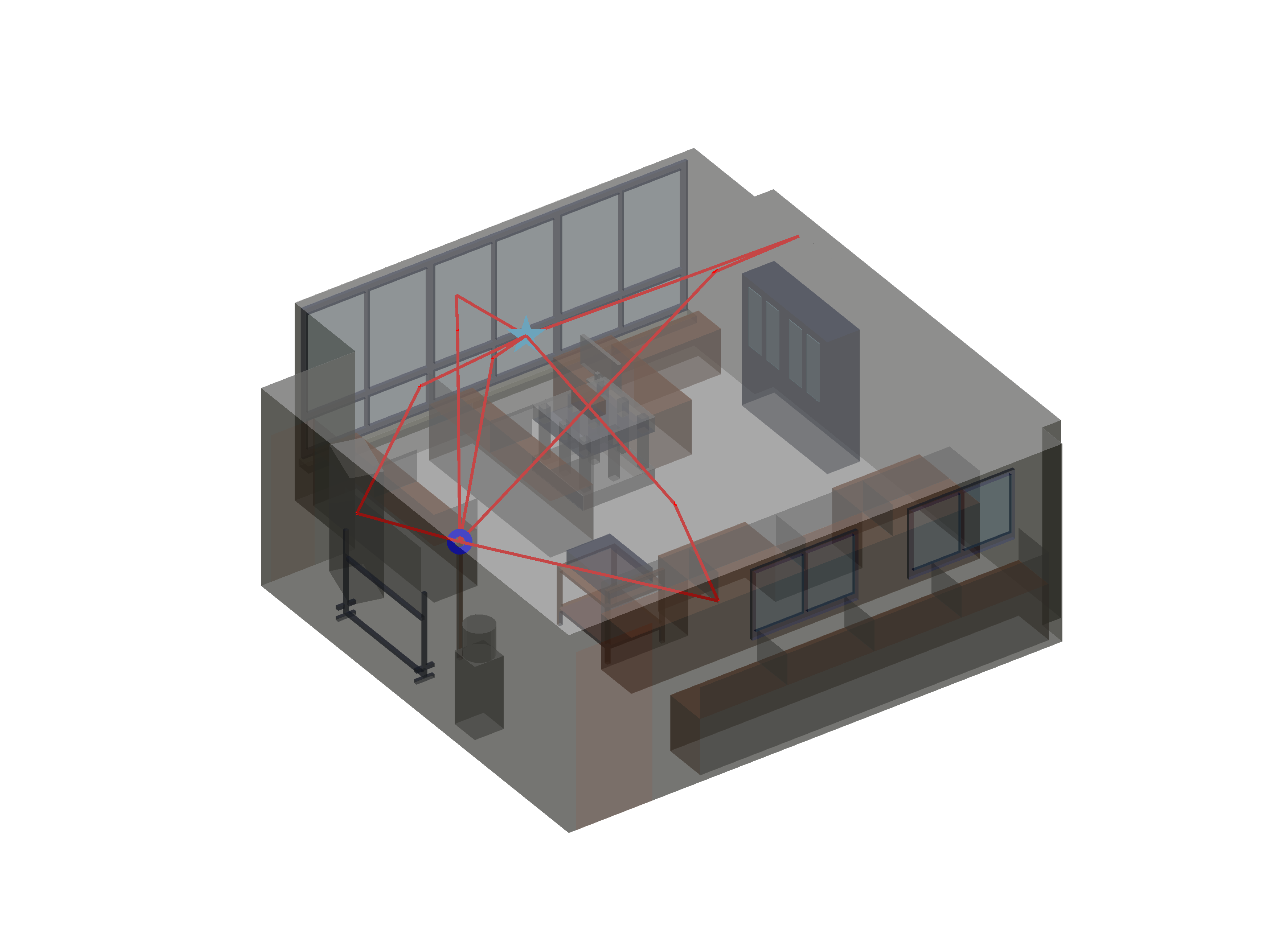}		
		\label{Fig6a}
	}
	\hspace{-4mm}
	\subfigure[]{
		\centering
		\includegraphics[scale=0.45,trim=120 0 140 0,clip]{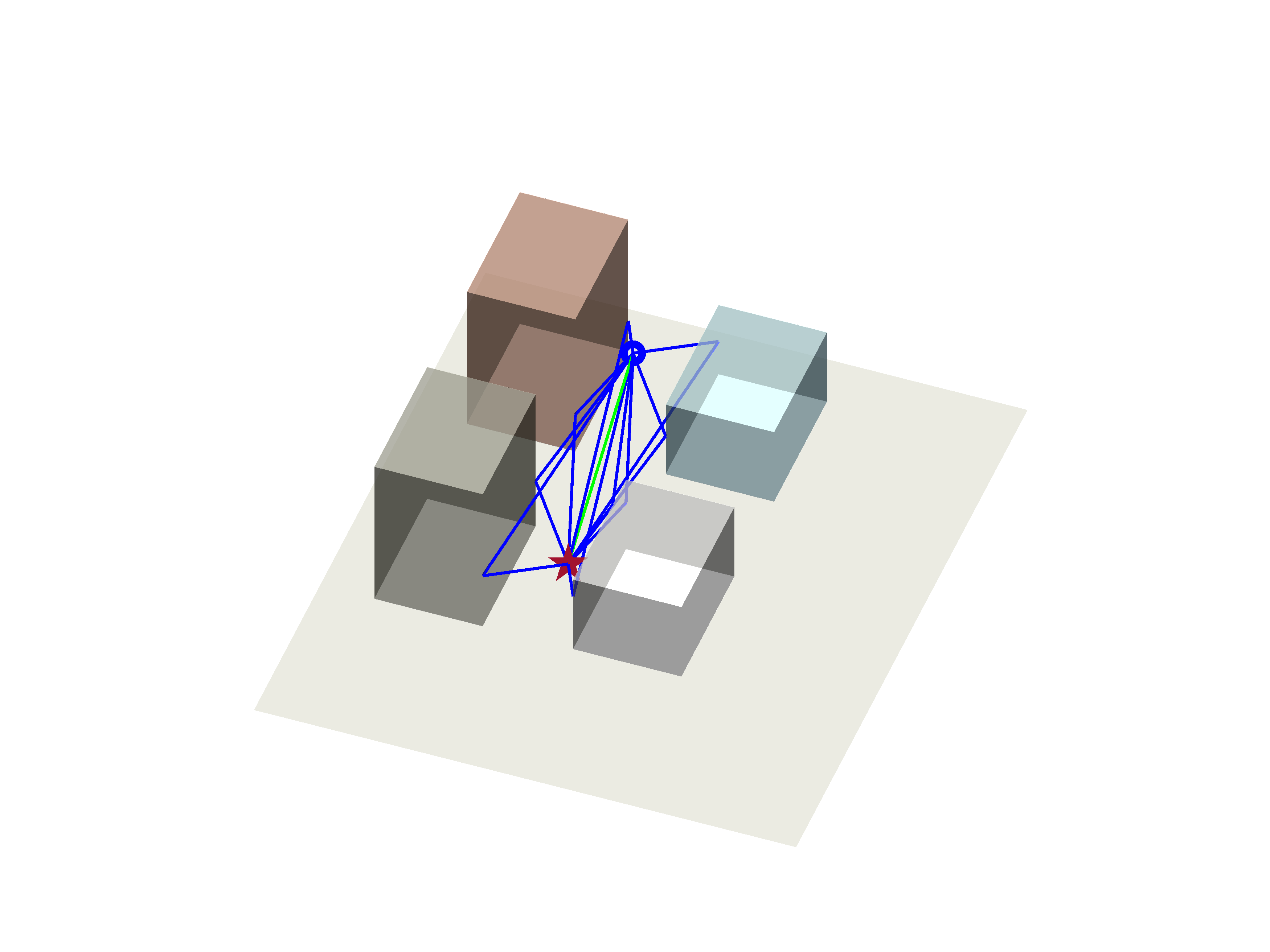}	
		\label{Fig6b}	
	}
	\setlength{\belowcaptionskip}{-0.5cm}
	
	\caption{(a) the office scene for reflection and (b) the outdoor scene for diffraction.}
	\label{Fig6}
\end{figure}

	An office scene in Fig.~\ref{Fig6a} is applied to verify the precise algorithm of reflection paths. This scene contains 1573 faces and 1183 wedges. The units are meters, with Tx at (4, 6.6, 1.6) and Rx at (1, 3.6, 1.6). The frequency is 2.4GHz. An outdoor scene in Fig.~\ref{Fig6b} is used to demonstrate the performance of the algorithm of diffraction paths, with Tx at (2, 7, 4), Rx at (20, 10, 2) and the units are also meters. The frequency is still 2.4GHz. Material information is included in the scene file. At the ray launching end, 4412 rays are launched. We allow the precise algorithm to run ten times on the inaccurate path. Comparisons are made among SBR with precise algorithm, IM \cite{ref4} and SBR in \cite{ref9}, referred to as traditional SBR.
	
\begin{figure}[tb]
		\centering
\vspace{0mm}	
\subfigure[]{
	\centering
	\includegraphics[scale=0.3,trim=10 0 30 10,clip]{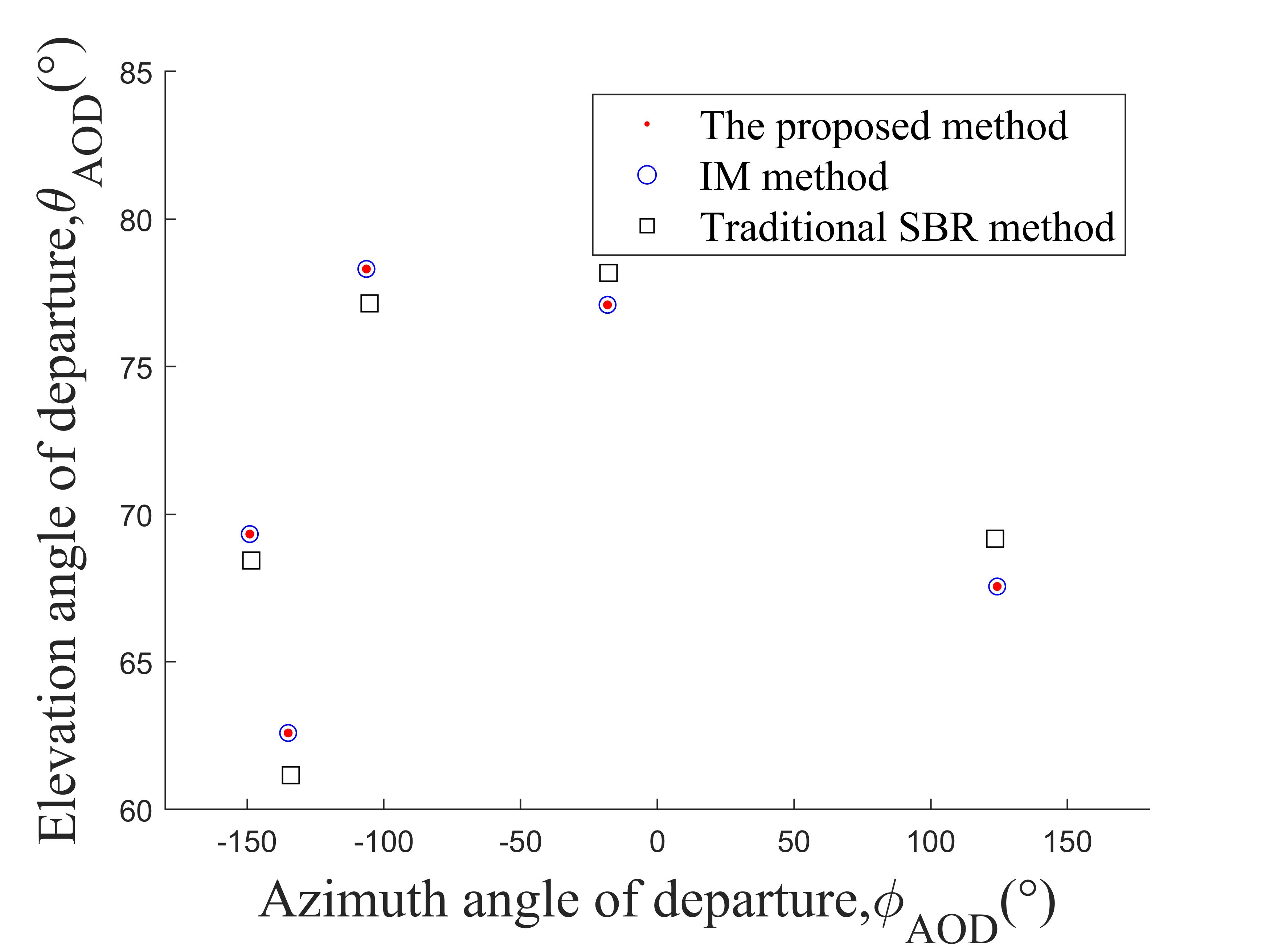}		
	\label{Fig7a}
}
\hspace{0mm}
\subfigure[]{
	\centering
	\includegraphics[scale=0.3,trim=10 0 30 10,clip]{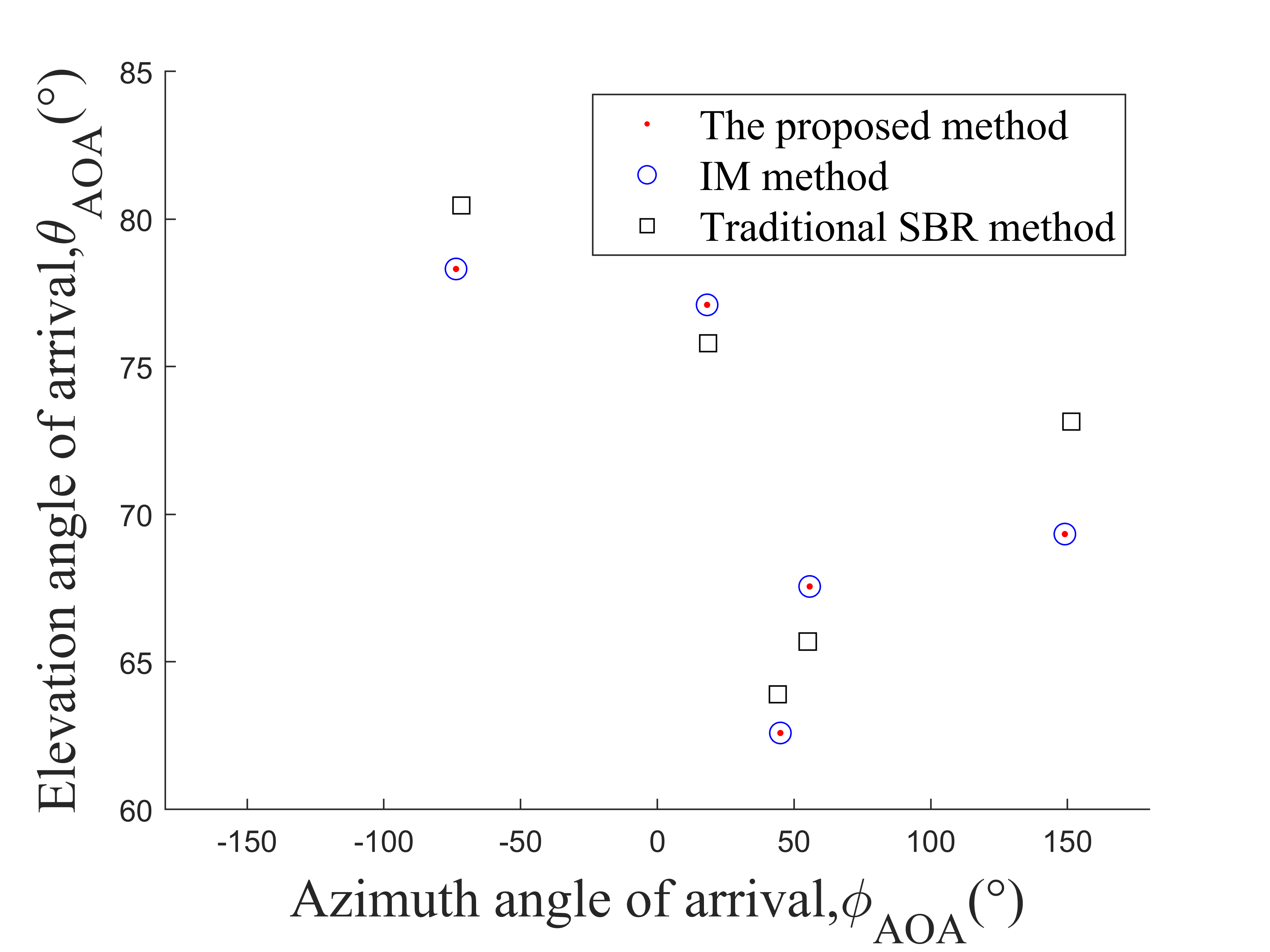}	
	\label{Fig7b}	
}
\qquad	
\vspace{-12mm}	
\subfigure[]{
	\centering
	\includegraphics[scale=0.3,trim=5 0 35 10,clip]{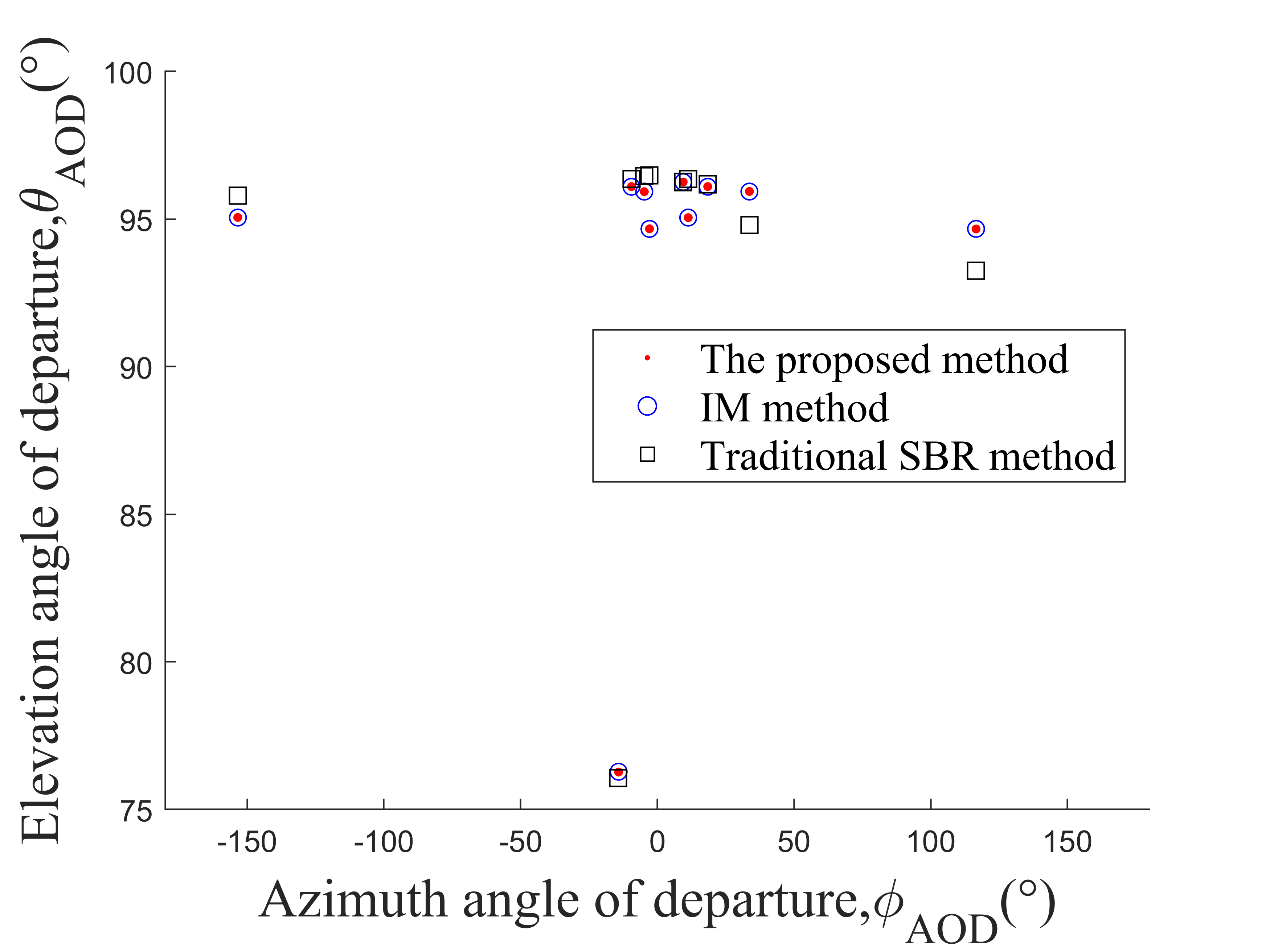}
	\label{Fig7c}		
}
\hspace{0mm}
\subfigure[]{
	\centering
	\includegraphics[scale=0.3,trim=5 0 35 10,clip]{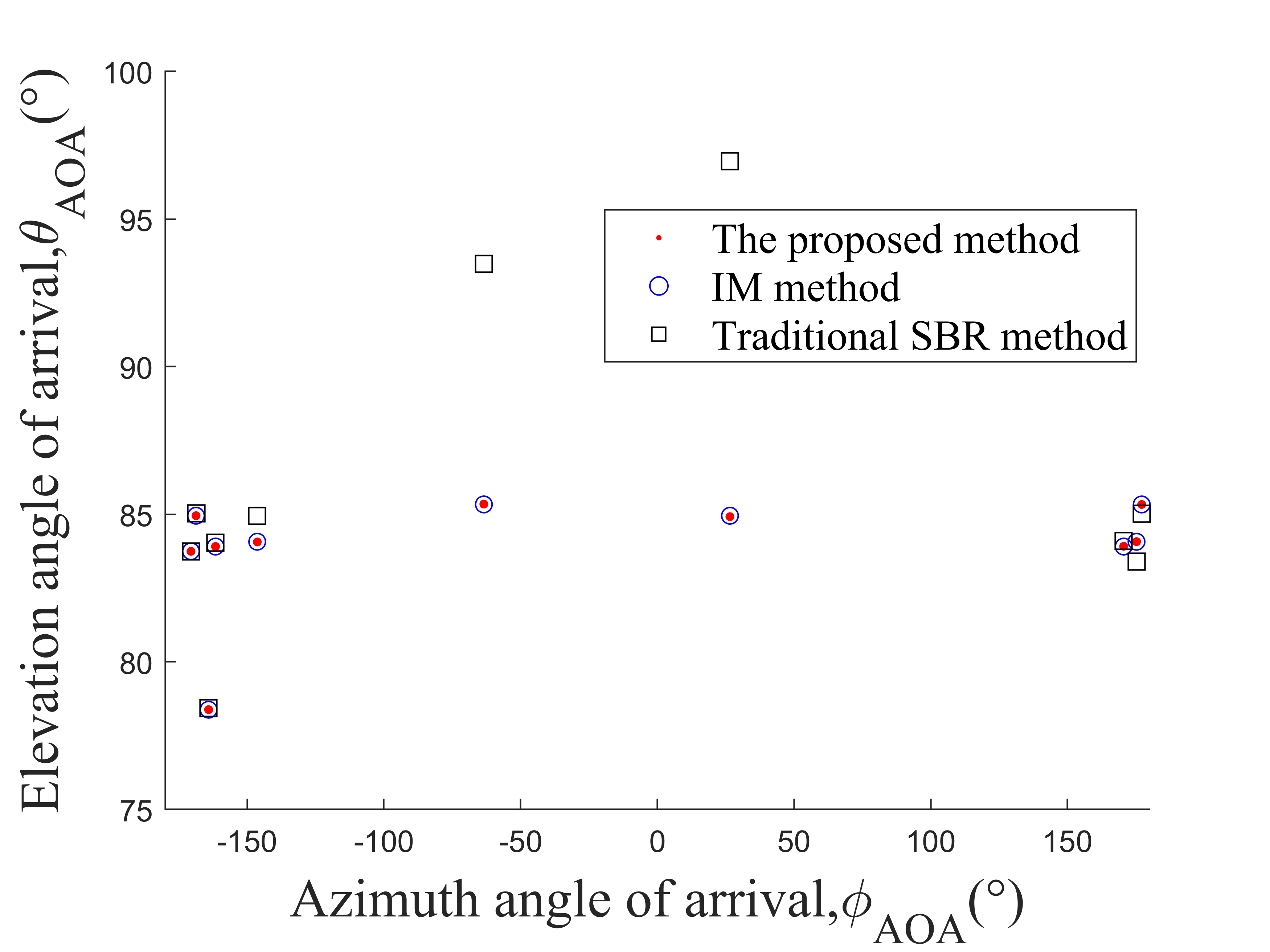}
	\label{Fig7d}		
	
}
\vspace{8mm}	
\setlength{\belowcaptionskip}{-0.5cm}

\caption{Results of traditional SBR, IM and SBR with precise algorithm of (a) AOD, (b) AOA of order 2 reflection and (c) AOD, (d) AOA of order 1 diffraction.}
\label{Fig7}
	\end{figure}

	Order 2 reflection paths are obtained in the office scene. After using the precise algorithm, SBR obtains five paths, with the same number of IM. Fig.~\ref{Fig7a} and Fig.~\ref{Fig7b} reveal the improvement of the angles. The average error of AOD of traditional SBR compared with IM is $1.4352^{\circ}$, and after 10 times of precise algorithm this error is reduced to $0.0062^{\circ}$. The error of AOA is always larger than that of AOD. Small error of AOD superimposes with each reflection or diffraction, resulting in a large error of AOA. So there is no numerical error limit for AOA. But with AOD error declining exponentially, error of AOA also reduces in this trend.
	
	Order 1 diffraction paths are also obtained based on the outdoor scene. The results of AOD and AOA of diffraction paths are shown inf Fig.~\ref{Fig7c} and Fig.~\ref{Fig7d}. The proposed method reduces AOA error from $2.2531^{\circ}$ to $0.0080^{\circ}$ after 10 iterations. This proves the proposed method can reduce angle error of SBR by at least two orders of magnitude.

	\begin{figure}[t]
		\centering
		\vspace{0mm}
		\subfigure[]{
			\centering
			\includegraphics[scale=0.3,trim=10 0 30 30,clip]{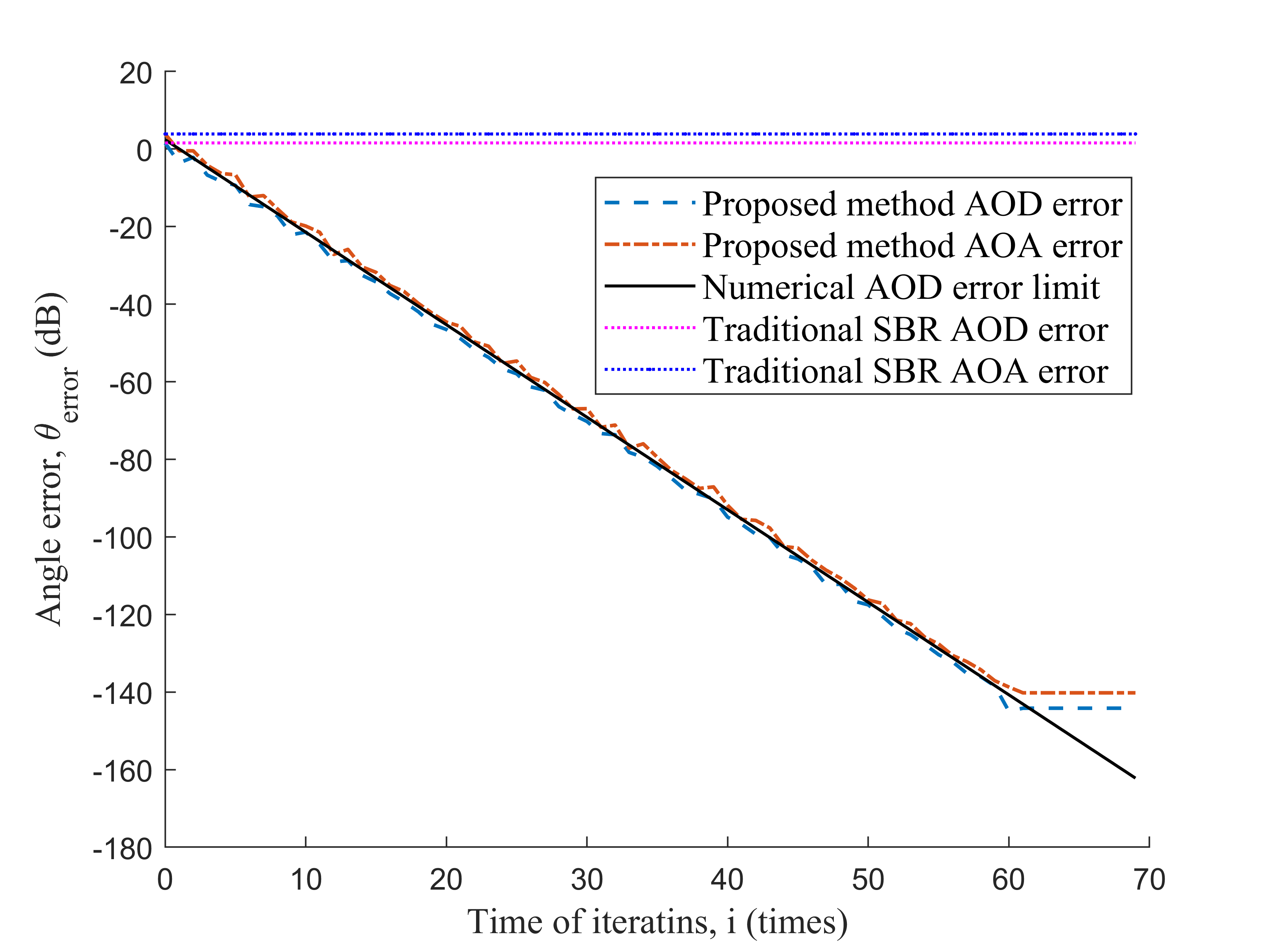}	
			\label{Fig8a}
	}	
		\subfigure[]{
	\centering
			\includegraphics[scale=0.3,trim=10 0 30 30,clip]{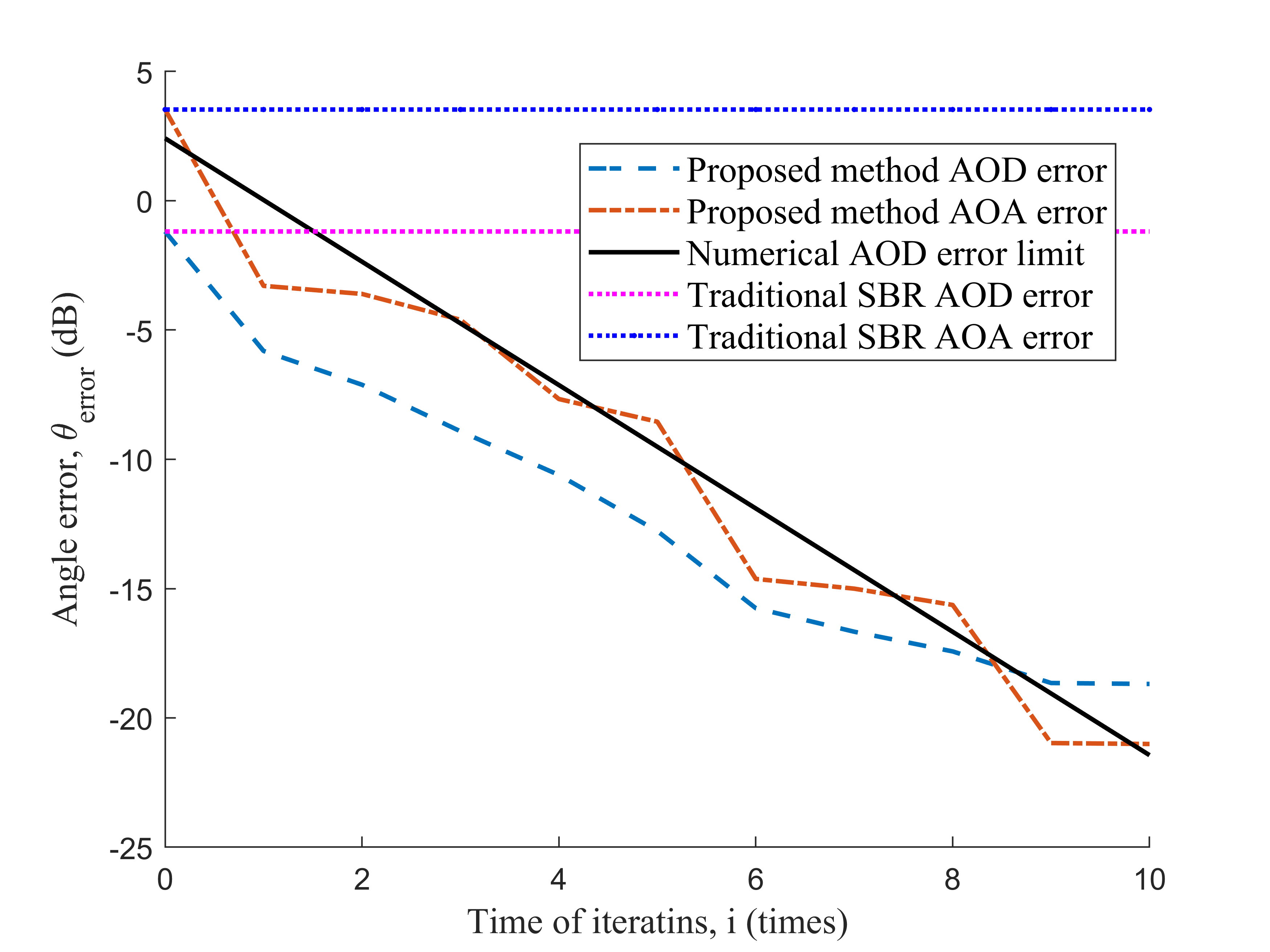}	
	\label{Fig8b}
}	
		\setlength{\belowcaptionskip}{-0.5cm}
		\caption{Relationship between angle error of AOD and AOA of reflections with number of times running iterations of (a) reflection scene and (b) diffraction scene.}
		\label{Fig8}
	\end{figure}

	Fig.~\ref{Fig8} shows the relationship between exact boost of angles per iteration and the number of iterations of both scenes. Angle error refers to the angle between the departure directions of a correct path and a SBR path. The error in degree is taken decibels, because of the exponential decline. The average AOD error of reflection scene in Fig.~\ref{Fig8a} is always below numerical limit of (\ref{eq10}) before 60 iterations, proving the proposed algorithm can improve accuracy exponentially. After iterating 60 times, error remains unchanged and is smaller than $10^{-14\circ}$. In the diffraction scene, error is reduced by 20 dB by 10 iterations, shown in Fig.~\ref{Fig8b}.

	In fact, there are some upturned points of mean AOD error in Fig.~\ref{Fig8a}, despite all points dropping from the former one. But this does not mean error, because six sub-ray cones can cover the original ray cone, so the sub-ray cone with a larger error contains the correct precise path. After one upturn, the error is significantly reduced. This kind of upturning is similar to the process of taking the optimal solution in the optimization algorithm, and the upturning enables this algorithm to avoid erring local optimal solutions.
		
\begin{figure}[t]
		\centering
\vspace{0mm}
\subfigure[]{
	\centering
	\includegraphics[scale=0.3,trim=10 0 30 20,clip]{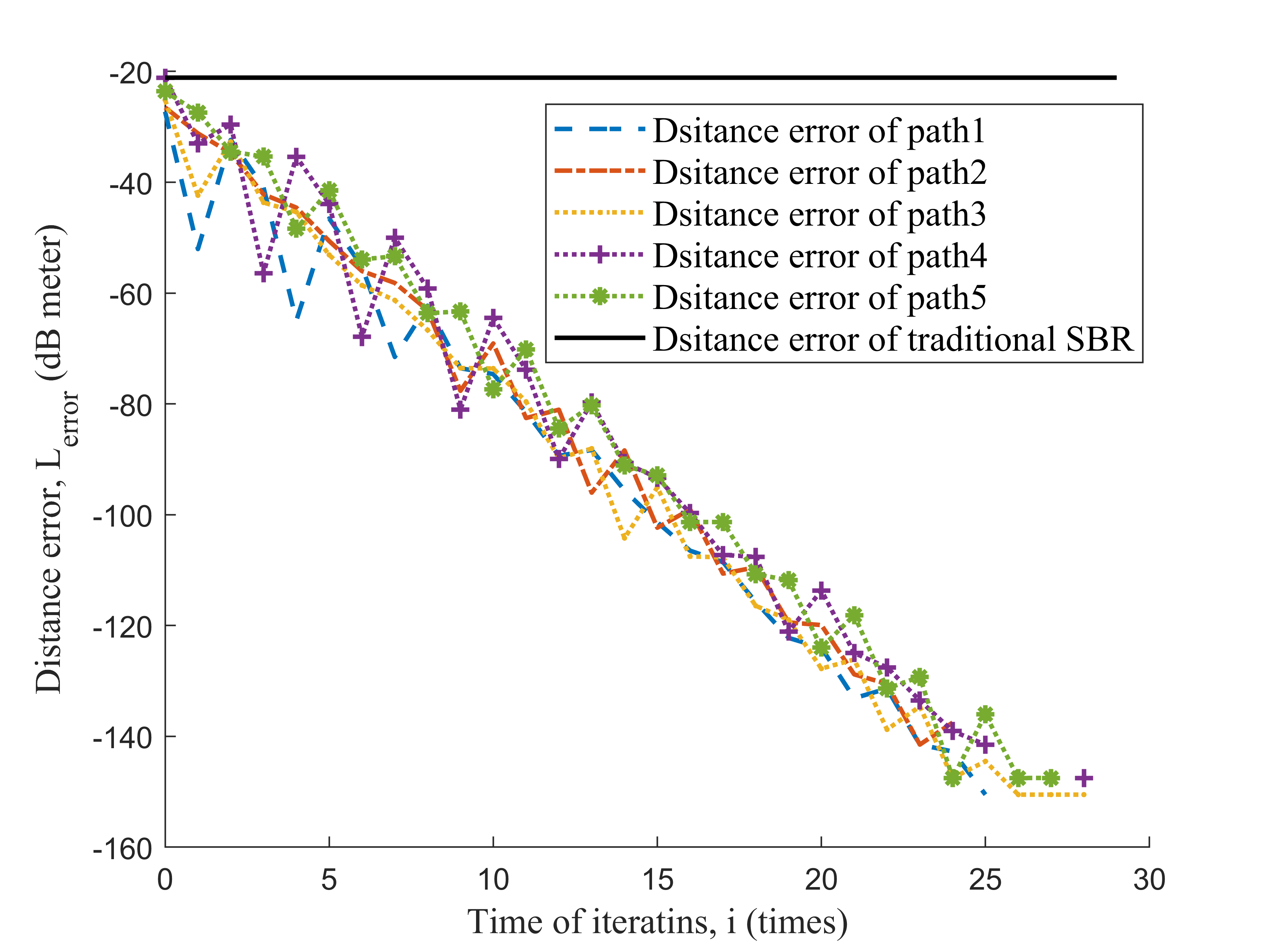}	
	\label{Fig9a}
}	
\subfigure[]{
	\centering
	\includegraphics[scale=0.3,trim=10 0 30 20,clip]{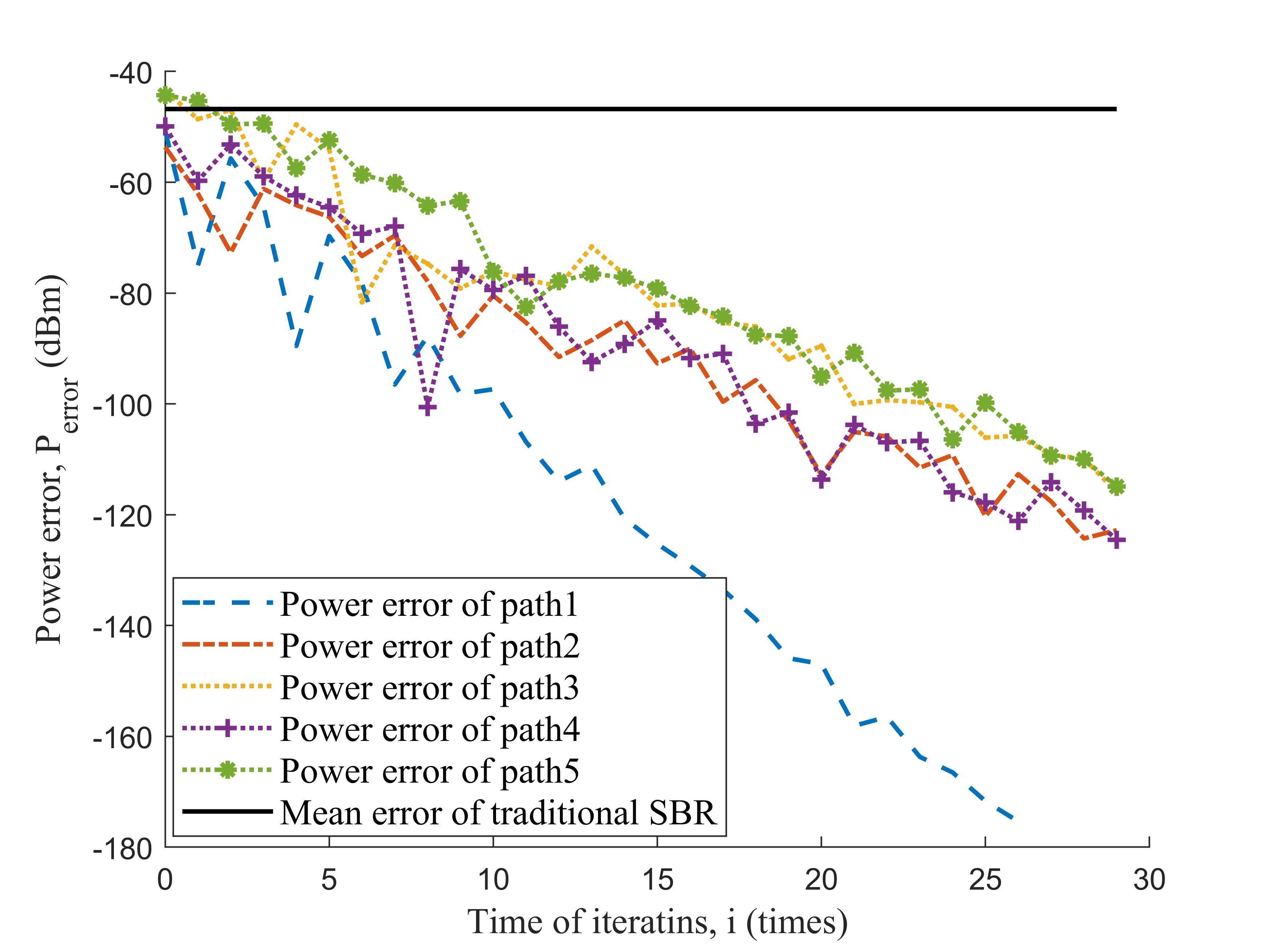}	
	\label{Fig9b}
}	
\setlength{\belowcaptionskip}{-0.5cm}
		\caption{Relationship between (a) distance error in decibel meters and itreations, and (b) power error and itreations.}
		\label{Fig9}
	\end{figure}

	Distance error is in Fig.~\ref{Fig9a}, which is the logarithmic difference between correct path length and obtained path length in meters. It also experiences exponential reduction using the precise algorithm. Power error is in Fig.~\ref{Fig9b}, representing the difference between obtained path power and precise results of IM, with the unit being dBm. Some points are lost when $i$ is close to 30, owing to the error reduced to zero, meaning the distances of all paths are exactly the same with those of IM. These two errors both follow the 10 iterations for 24dB error reduction law, which is derived from (\ref{eq10}).

\begin{figure}[t]
		\centering
		\vspace{0mm}
		\includegraphics[scale=0.4,trim=10 0 30 10,clip]{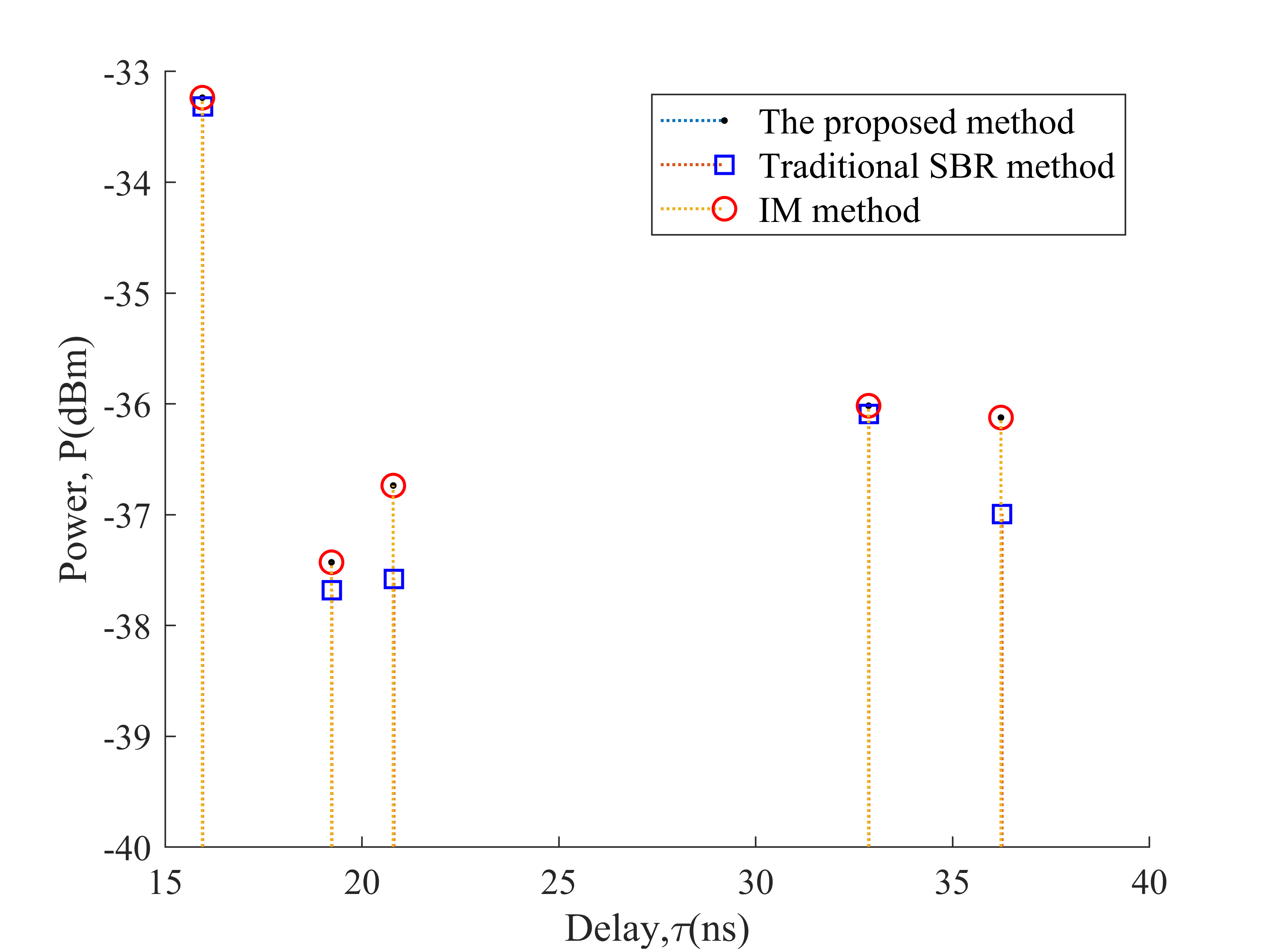}		
		\setlength{\belowcaptionskip}{-0.5cm}
		\caption{PDP of precise algorithm, traditional SBR and IM.}
		\label{Fig10}
	\end{figure}

	PDP shows the delay and power of paths. As shown in Fig.~\ref{Fig10}, traditional SBR has some errors compared with IM, while the precise algorithm gets almost the same results.

	Finally, the time cost is compared between the methods in Table~\ref{Tab2}. Because this algorithm is based on paths, the time cost of it is mostly related to the number of paths and times of iterations despite the number of faces and wedges. The time cost of proposed method is little cmopared with SBR and IM.

	\begin{table}[t]
		\centering
		\vspace{2mm}
		\caption{Time cost of three methods.}
		\setlength{\tabcolsep}{6pt} 
		\renewcommand\arraystretch{1.6} 
		\label{Tab2}
		\begin{tabular}{|l|p{1.5cm}|p{1.5cm}|p{1.5cm}|}
			\hline
			Order & IM time (s) & Traditional SBR time (s) & Precise time (s) \\
			\hline
			2 reflections & 15.0420 & 4.4633 & 0.9273 \\
			\hline
			3 reflections & 1196.6591 & 10.2837 & 3.8437 \\
			\hline
		\end{tabular}
	\vspace{-2mm}
	\end{table}

	\section{Conclusions}
	
	In this paper, an iterative precise SBR algorithm based on equiangular division has been proposed. Errors of AOD, AOA, length and power of paths can be reduced to any given level with the increase of iteration numbers. Typically, 10 iterations can lead to 24dB reduction in error of all parameters above.	Angle distribution, path length and PDPs have been compared to evaluate the improvement. The time cost of the algorithm is little compared with traditional SBR, and much small compared with IM. In diffraction scenes, although the error of AOD and AOA declines rapidly, it remains around 0.01m after ten iterations. Though this result is much more accurate than traditional SBR, there can be some further research on this phenomenon. Also, the proposed method can be applied to MIMO systems in the future.
	
	\section*{Acknowledgment}
	This work was supported by the National Key R\&D Program of China under Grant 2018YFB1801101, the National Natural Science Foundation of China (NSFC) under Grants 61960206006 and 61901109, the Frontiers Science Center for Mobile Information Communication and Security, the High Level Innovation and Entrepreneurial Research Team Program in Jiangsu, the High Level Innovation and Entrepreneurial Talent Introduction Program in Jiangsu, the Research Fund of National Mobile Communications Research Laboratory, Southeast University, under Grant 2021B02, and the EU H2020 RISE TESTBED2 project under Grant 872172.
	\small {}

	\bibliographystyle{IEEEtran}
	
\end{document}